\documentstyle[twocolumn,epsf,aps]{revtex}
\begin{document}
\newcommand{\beq}{\begin{equation}}
\newcommand{\eeq}{\end{equation}}

\title{Master equation approach to protein folding}
\author{
Marek Cieplak$^1$, Malte Henkel$^2$, 
and Jayanth  R. Banavar$^3$}
\address{$^1$ Institute of Physics, Polish Academy of Sciences,
and College of Sciences, 02-668 Warsaw, Poland}
\address{$^2$ Laboratoire de Physique des Mat{\'e}riaux, Universit{\'e}
Henri Poincar{\'e} Nancy I, F-54506 Vand{\oe}uvre, France} 
\address{$^3$ Department of Physics and Center for Materials Physics,
104 Davey Laboratory, The Pennsylvania State University, 
University Park, PA 16802}

\address{
\centering{
\medskip\em
{}~\\
\begin{minipage}{14cm}
%%{}~~~ 
The dynamics of two 12-monomer heteropolymers on the square lattice
is studied exactly within the master equation approach.
The time evolution of the
occupancy of the native state is determined.
At low temperatures,
the median folding time follows the Arrhenius law and
is governed by the longest relaxation time.
For both good and bad folders, 
significant kinetic traps 
appear in the folding funnel and the kinetics 
of the two kinds of folders are quite similar.
What distinguishes between the good and bad folders are the
differences in their thermodynamic stabilities.
{}~\\
{}~\\
{\noindent PACS numbers: 87.15.By, 87.10.+e}
\end{minipage}
}}

\maketitle

A denatured protein folds into a compact native state in a time of order
a millisecond after the physiological conditions are restored \cite{1}.
This process is reversible and the native state is believed
to coincide with the ground state of the system.
Studies of simplified lattice models of proteins have provided many
insights into the dynamics of the folding process \cite{2}.
The crucial feature that makes lattice models 
useful is the possibility of enumerating
the complete set of conformations and determining
which of them is the ground state. This can be accomplished, if
the length of the heteropolymer, $N$, is small.
Such studies have elucidated the role
of thermodynamic stability\cite{2,3}, stability against 
mutations \cite{4}
and existence of a linkage between rapid folding and the stability
of the native state\cite{3}.

The key concept that has been introduced to explain the
rapid folding that occurs in natural proteins is that of the
folding funnel \cite{5,6} -- a set of conformations that
are smoothly connected to the native state, as indicated 
schematically at the top of Figure 1. It is expected that for
random sequences of aminoacids
there are competing basins of attraction which would trap the system
away from the native state. This corresponds schematically
to what is shown at the bottom of Figure 1.
The identification of states that belong to the funnel is not easy
due to an enormous number of conformations 
that are present even for a small $N$  and because the
problem is dynamical one. Possible approaches 
include the monitoring of frequences of passages between
various states in Monte Carlo trajectories \cite{6}
or the mapping of states into underlying valleys of
effective states \cite{7,8}.

%%++++++++++++++++++++++++++++++++++++++++++++++++++++++++++++++++++++++++++++++
%FIGURE 1
\begin{figure}
\epsfxsize=3.2in
\centerline{\epsffile{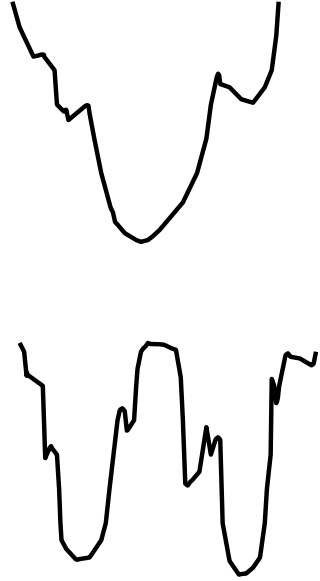}}
\caption{
A schematic representation of the phase space properties
of good (top) and bad (bottom) folders. The vertical axis corresponds
to energy  and the horizontal axis to a "coordinate" in the phase space.
}
\end{figure}
%%++++++++++++++++++++++++++++++++++++++++++++++++++++++++++++++++++++++++++++++

Essentially all approaches to the studies of the folding dynamics
have been restricted to Monte Carlo simulations that start from a few
randomly chosen initial conformations\cite{9}. 
%These Monte Carlo
%studies typically do not implement the detailed balance condition.
The only exception has been an approach due to Chan and Dill \cite{10}
in which an enumeration of 
transition rates between classes of 
conformations which have the same number of contacts and are
a given number of kinetic steps away from the native state.

Recently, we presented an exact method to study the dynamics of
short model proteins \cite{11} which was
based on the master equation\cite{12}. 
We have specifically considered two $N$=12 sequences, called A and B,
which are placed on a square lattice.
The dynamics of these sequences can be studied exactly
because the sequences can acquire only ${\cal N}$=15037 conformations.
The two sequences have the same set of contact
energies $B_{ij}$ but their assignment to various monomers
$i$ and $j$ is different with the
result that A is a good folder and B is a bad folder.
The basic finding of \cite{10} is that the qualitative
picture corresponding to Figure 1 is indeed correct.
By identifying kinetic trap states that are responsible
for the slowest dynamical processes in the system we could
demonstrate that the traps for sequence A are within the folding funnel
whereas for sequence B the relevant traps form a valley
which competes with the native valley in analogy to the bottom
of Figure 1. In the current paper we provide a deeper characterization
and comparisons of the two sequences and 
elucidate the nature of the trap states.

%%++++++++++++++++++++++++++++++++++++++++++++++++++++++++++++++++++++++++++++++
%FIGURE 2
\begin{figure}
\epsfxsize=3.2in
\centerline{\epsffile{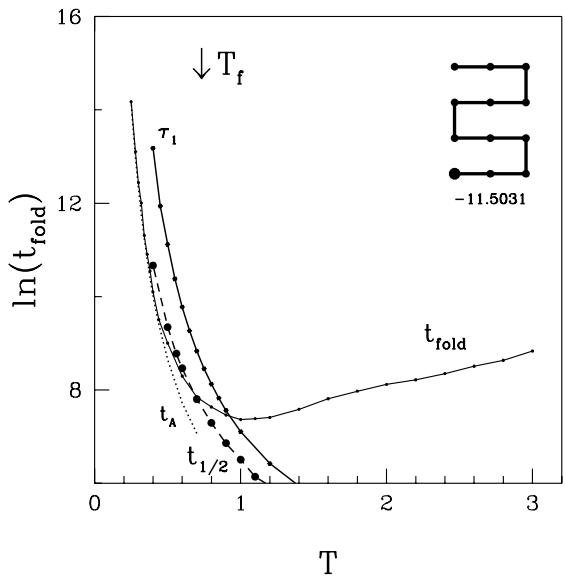}}
\caption{
Top: The native conformation 
and its energy for sequence A. The enlarged circle shows the first
monomer.
Main: Dynamical data for the folding.
The solid line marked by $t_{\rm fold}$ gives the median folding time
derived from  1000 Monte Carlo trajectories.
The solid line $\tau _1$ is the longest relaxation time.
The broken line $t_{\frac{1}{2}}$ with the black circles  gives the
time for $P_0(t)$ to reach $\frac{1}{2} P_0$ 
from the uniformly random initial state. 
The dotted line $t_A$ is a fit of the Monte Carlo data
to the Arrhenius law with $\delta E$=2.76. The arrow at the
top indicates the value of the folding transition temperature.
}
\end{figure}
%%++++++++++++++++++++++++++++++++++++++++++++++++++++++++++++++++++++++++++++++
The sequences that we study are described by the Hamiltonian
\begin{equation} \label{eq:equat1}
{\cal H} = \sum_{ij} B_{ij}  \Delta _{ij} \;\;,
\end{equation} 
where $\Delta _{ij}$ indicates that the
contact interaction $B_{ij}$ is assigned to monomers
which are geometrical nearest neighbors on the lattice
but are not neighbors along the sequence. In such arrangements
$\Delta _{ij}$ is equal to 1, otherwise it is 0. 
The values of the 25 
couplings are chosen as Gaussian numbers which are centered
around -1 to provide an overall attraction, as detailed in \cite{10}.
The ground state of sequence A is maximally compact and it fills the
$3 \times 4$ lattice, as shown at the top of Figure 2. 
For sequence B, the ground state, shown at the top of Figure 3, is
doubly degenerate. Both of the ground states are compact but not
maximally compact. They differ merely by a placement of one end monomer
and therefore they were considered as an effective single
native state.

%%++++++++++++++++++++++++++++++++++++++++++++++++++++++++++++++++++++++++++++++
%FIGURE 3
\begin{figure}
\epsfxsize=3.2in
\centerline{\epsffile{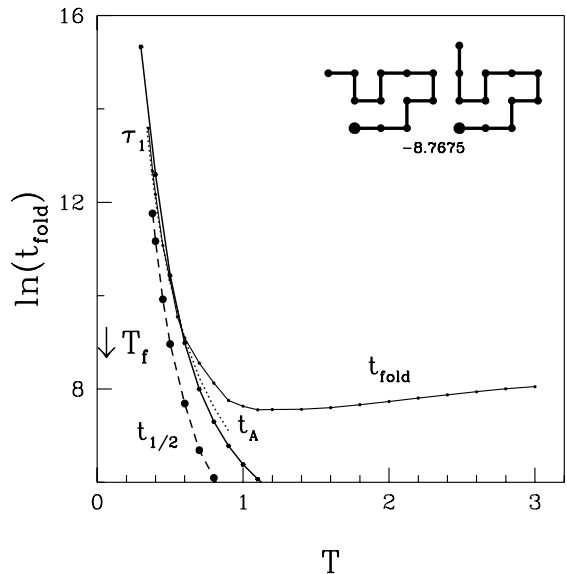}}
\caption{
Same as Figure 2, but for sequence B. 
For the curve $t_A$, $\delta E$=3.55. There are two native
conformations
with the same energy.
}
\end{figure}
%%++++++++++++++++++++++++++++++++++++++++++++++++++++++++++++++++++++++++++++++

We have found that the dynamics of A and B
are superficially similar:  for both, the median 
folding time, $t_{\rm fold}$, and
the longest relaxation time, $\tau _1$ diverge  at low $T$ according
to an Arrhenius law.  The temperature $T_{\rm min}$ at which
folding to the native state proceeds the fastest is
about the same for both sequences.
It is the location of the folding transition 
temperature, $T_{\rm f}$, with respect to 
$T_{\rm min}$ which distinguishes between sequences A and B. 
$T_{\rm f}$ is defined as the temperature at which the equilibrium
value of the probability to occupy the native state, $P_0$, crosses 
$\frac{1}{2}$  and is a measure of thermodynamic stability. 
For bad folders,
$T_{\rm f}$ is well below $T_{\rm min}$ and thus a substantial occupation
probability for  
the native state is found only in a temperature range in which the
dynamics are glassy. 
For sequence A, the values of 
$T_{\rm f}$ and $T_{\rm min}$ are 0.71 and $1.0\pm 0.1$, 
respectively, while for
sequence B, the corresponding values are 0.01 and $1.1\pm 0.1$. 
Thus the two sequences are dynamically similar
but the equilibrium properties differ dramatically.

\vspace{0.5cm}

\noindent
{\bf Methods:$\;\;$}
We study our sequences  through an analysis of the master equation
and then compare the results to those obtained by the
Monte Carlo approach.
The master equation does not deal with a specific trajectory
but with an ensemble of trajectories and it reads
\begin{equation}
\frac{\partial P_{\alpha}}{\partial t} = \sum_{\beta \neq \alpha}
\left[ w(\beta \to \alpha) P_{\beta} - w(\alpha \to \beta) P_{\alpha} \right]
\end{equation}
where $P_{\alpha} = P_{\alpha}(t)$ is the probability of finding the 
sequence in conformation $\alpha$ at time $t$. The quantity 
$w_{\alpha \beta}=w(\beta \to \alpha)$ is the transition rate 
from conformation
$\beta$ to conformation $\alpha$. 
Of course, writing a master equation relies 
on the assumption that a Markov chain description adequately
describes protein kinetics, in particular, memory effects are assumed to 
be negligible. The actual derivation of a master equation 
remains a formidable problem in itself \cite{Kreu81}. 

One may bring this into a matrix form by
letting $\vec{P} = (P_1,\ldots,P_{\cal N})$ and 
\begin{equation} \label{StochBed}
m_{\alpha \beta } = - w_{\alpha \beta} \leq 0 \;\;
 \mbox{\rm if $\alpha \neq \beta$} \;\; , \;\;
m_{\alpha \alpha} = \sum_{\beta \neq \alpha} w_{\beta \alpha} \;\; .
\end{equation}
The master equation then takes the form of an imaginary-time
Schr\"odinger equation \cite{12,13} 
\begin{equation}
\partial_t \vec{P} = - \hat{M} \vec{P} \;\;,
\end{equation}
where the $m_{\alpha \beta}$ are the matrix elements of $\hat{M}$. 

The time-dependence state vector $\vec{P}(t)$ at time $t=n\tau _0$ can
be obtained by applying $n$ times the recursion $\vec{P}((n+1)\tau _0)=
(1 - \tau _0 \hat{M}) \vec{P}(n\tau _0)$ to $\vec{P}_{\rm in}$,
where $\tau _0$ is a microscopic time associated with a single move
and $\vec{P}_{\rm in}$ is some initial probability distribution.
On the other hand, the spectrum of relaxation times 
$\tau_{\lambda} = 1/(\mbox{\rm Re } M_{\lambda}) \geq 0$
follows directly from the eigenvalues $M_{\lambda}$ of $\hat{M}$. 

The transition rates $w_{\alpha \beta}$ are designed so that
a stationary solution of the master equation corresponds to 
equilibrium with the Hamiltonian given by eq.~(\ref{eq:equat1}).
This can be accomplished provided
the detailed balance condition
\begin{equation} \label{DetGG} 
w_{\alpha \beta} P_{\beta}^{\rm eq} = 
w_{\beta \alpha} P_{\alpha}^{\rm eq}
\end{equation}
is satisfied. Here, $P_{\alpha}^{\rm eq} \sim e^{-{E\alpha }/T}$
is a stationary solution of the master equation and $E_{\alpha}$
is the energy of the sequence in conformation $\alpha$.
Eq. (\ref{DetGG}) is satisfied by any
$w_{\alpha \beta}=f_{\alpha \beta}
 \exp \left[ -(E_{\alpha} - E_{\beta}) /2T \right]$ 
provided only that $f_{\alpha \beta}=f_{\beta \alpha}$.
Here, we choose $w_{\alpha \beta}=w_{\alpha \beta}^{(1)}+
w_{\alpha \beta}^{(2)}$, where 
\begin{equation} \label{eq:Raten}
w_{\alpha \beta}^{(\sigma)} = \frac{2}{\tau _0} R_{\sigma} 
\cosh\left[ \left( E_{\alpha} - E_{\beta}\right) /T\right]
\end{equation}
with $R_1 + R_2 =1$. Here, $\sigma= 1$ and 2 refer to 
the single- and double-monomer moves respectively.
It is understood that $w_{\alpha \beta}^{(\sigma)}=0$ if there is 
no move of
type $\sigma$ linking $\beta$ with $\alpha$. 
The choice (\ref{eq:Raten}) guarantees that transition
rates are finite and bounded for all temperatures. In analogy to
\cite{3}, we focus on $R_1=0.2$ and take the single and two-monomer
(crankshaft) moves as in \cite{10}.

Because of the detailed balance condition, 
the eigenvalues $E_{\alpha}$ are not calculated by
diagonalizing $\hat{M}$ directly, but 
by diagonalizing an auxiliary matrix $\hat{K}$ with
elements
\begin{equation}
k_{\alpha \beta} = \frac{v_{\beta}}{v_{\alpha}} m_{\alpha \beta} =
k_{\beta \alpha} \;\;,
\end{equation}
where $v_{\alpha}= e^{-E_{\alpha}/2T}$. 
The master equation can be then rewritten as
\begin{equation}
\partial_t \vec{Q} = - \hat{K} \vec{Q} \;\;,
\end{equation}
where $\hat{K}$ is real symmetric and $Q_{\alpha}=P_{\alpha}/v_{\alpha}$.
This equation is then diagonalized using the standard symmetric
Lanczos algorithm without reorthogonalization, as described in 
\cite{14,15} and then the eigenmodes of $\hat{M}$ are determined.
In particular, since $\hat{K}$ and $\hat{M}$ are related by a similarity
transformation, all eigenmodes $M_{\lambda}$ are real. 

The eigenvector
corresponding to $M_{\lambda}=0$, i.e. to the infinite 
relaxation time, determines the equilibrium occupancies of the
conformations. This eigenvector needs to be normalized to 
satisfy its
probabilistic interpretation. 
The longest finite relaxation time 
$\tau _1 =1/M_1$ is found from the smallest non-zero eigenvalue $M_1$.

Our Monte Carlo simulations have been done in a way that satisfies the
detailed balance conditions and were devised along the
lines described in \cite{10}. For polymers, satisfying these
conditions is non-trivial because each conformation has its own
number, $K$,  of allowed moves that the conformation can make. Thus
the propensities to make a move in a unit time vary from conformation
to conformation and the effective "activities" of the conformations
need to be matched. This can be accomplished by first determining the
maximum value of $K$, $K_{\rm max}$.  For the 12-monomer chain on the
square lattice, $K_{\rm max}$ is equal to 14.
We associate a single time
unit with the conformations in which $K$=$K_{\rm max}$.   This means that
each allowed move is being attempted always with probability
$1/K_{\rm max}$. For a conformation with $K$ allowed moves, the probability
to attempt any move is then $K/K_{\rm max}$ and the probability not to 
carry out any
attempt is $1\;-\;K/K_{\rm max}$.  The attempted moves are then accepted or
rejected as in the standard Metropolis procedure.  
The probability of a single monomer move (an end flip or 
a switch in a pair of bonds that make the $90^o$ angle)
is additionally reduced by the factor of 0.2 and an
allowed double monomer crankshaft move by 0.8.  
The time is then measured
in terms of the total number of the Monte Carlo attempts divided by
$K_{\rm max}$.  This  scheme not only establishes the detailed balance
conditions but it also uses less CPU compared to a
process in which moves are attempted without regard to whether they
are allowed or not.

\vspace{0.5cm}

\noindent
{\bf Results:$\;\;$}
Figure 4 shows the 4 longest finite relaxation times for sequence A
as a function of temperature. It is seen that
they are all roughly proportional to each other
and they all appear to diverge at $T$=0, albeit with differing
energy barriers. The longest
relaxation time follows the Arrhenius law, $\tau _1\; \sim \; 
e^{{\delta E_1}/T}$ with $\delta E_1$ of about 4.1. The Monte Carlo
derived median folding time also follows the Arrhenius law
at low temperatures but with $\delta E = 2.76 \pm 0.06$. 
The overall Arrhenius-like behavior suggests that
the physics of folding at low  $T$'s is dominated by
processes that establish equilibrium. Their longest time scale
is then set by $\tau _1$. At temperatures above $T_{\rm min}$, 
$t_{\rm fold}$ no longer follows $\tau _1$.
Here, the physics of folding is  dominated by the statistics
of rare events: the system establishes equilibrium rapidly
and then folding is reached by a search in equilibrium.
The search becomes more and more random when $T$  becomes
larger and larger. For sequence B, similar results are obtained
but the Arrhenius barrier in $t_{\rm fold}$ is $3.55 \pm 0.06$,
whereas $\delta E_1$ = 4.0$\pm$0.06.
Figures 2 and 3 show fits to the Arrhenius law for sequences
A and B respectively.

%%++++++++++++++++++++++++++++++++++++++++++++++++++++++++++++++++++++++++++++++
%FIGURE 4
\begin{figure}
\epsfxsize=3.2in
\centerline{\epsffile{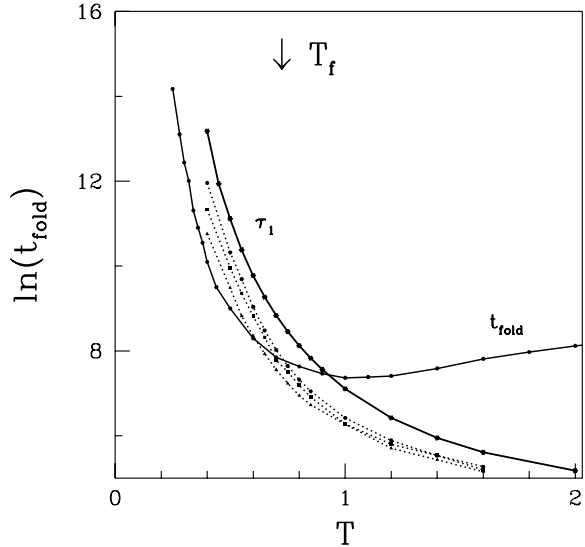}}
\caption{
The four longest relaxation times versus $T$ for sequence A.
The very longest, $\tau _1$, is drawn by a solid line
whereas the remaining three by the dotted lines.
The value of $T_{\rm f}$ is indicated by an arrow and the Monte Carlo data
on $t_{\rm fold}$ are repeated from Figure 2 for a comparison.
}
\end{figure}
%%++++++++++++++++++++++++++++++++++++++++++++++++++++++++++++++++++++++++++++++
 
We now focus on the time evolution of the probability, $P_0(t)$, to
occupy the native state, where $t$ denotes time. This probability
depends on the initial conditions. The solid lines in Figure 5 show
the evolution of $P_0(t)$ when
in the initial state 
all conformations have  equal probability of $1/{\cal N}$ to
be occupied. The main figure is for sequence A and 
the time evolution is shown for three temperatures: 1.2, 0.9 and 0.6.
The data for the lowest of these temperatures are also compared 
with the Monte Carlo evolution. The Monte Carlo data,
obtained by starting from random conformation, agree with the
exact evolution. 
The time scales at which the equilibrium
saturation values of $P_0$ are reached are of order $\tau _1$.
The top portion of Figure 5 shows $P_0(t)$ for sequence B.
It is seen that the saturation levels are significantly lower
for B than for A at the same temperatures. This reflects
a significantly lower value of $T_{\rm f}$ for B.
We also observe, by looking at Figures 2 and 3 that the median
folding time at low temperatures appears to coincide with the time
needed for $P_0(t)$ to reach half of its equilibrium value.

%%++++++++++++++++++++++++++++++++++++++++++++++++++++++++++++++++++++++++++++++
%FIGURE 5
\begin{figure}
\epsfxsize=3.2in
\centerline{\epsffile{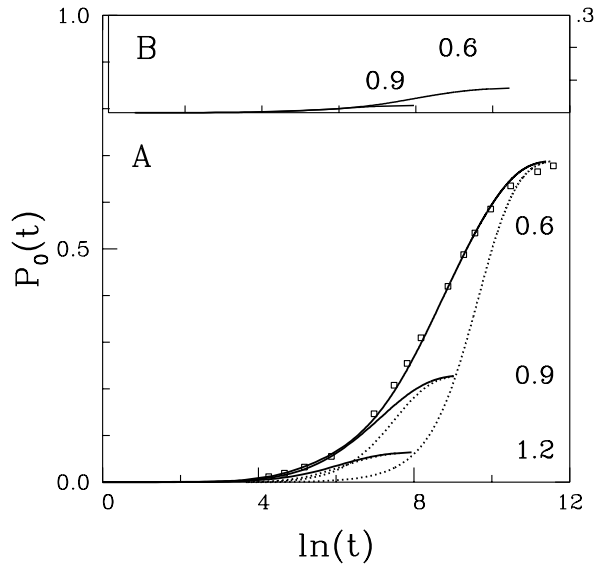}}
\caption{
Probability of occupation of the native state, $P_0(t)$,
of sequence A, 
for three values of the temperatures, indicated on the right.
$P_0(\infty)$ agrees with the equilibrium value.
The solid lines correspond to the uniformly random initial
condition and the dotted lines correspond to an initial
placement of the system in the trap state.
The squares correspond to Monte Carlo results, based on
200 random starting conformations.
The top figure shows $P_0(t)$ for sequence B.
The initial condition is that of uniform randomness.
The $y$-axis scale is indicated on the right.
}
\end{figure}
%%++++++++++++++++++++++++++++++++++++++++++++++++++++++++++++++++++++++++++++++

The dotted line in Figure 5 corresponds to the initial
state being the kinetic ``trap'' conformation \cite{16}
which is the strongest obstacle in reaching equilibrium.
If the system starts in the trap state, it will still reach
the final equilibrium in a time scale set by $\tau _1$ but
at shorter times the occupancy of the native state is significantly
less than when starting from a totally random state.

In general, it is not easy to determine which conformation
forms the most potent trap. One may attempt to determine it
by monitoring occupation of the local energy minima
in a Monte Carlo process performed at very low temperatures.
This approach has been successfully applied in \cite{8}.
Here, the trap state is determined in an exact way
by studying the eigenvector corresponding to the longest relaxation 
time and by identifying the local energy minima
which contribute the most to this eigenvector
at low $T$. The largest weight is associated with the native state.
Usually, the second largest corresponds to the most relevant trap.
This happens provided
the second largest occupancy is in a local
energy minimum state. A non-minimum state, immediately preceding
the native state kinetically, may also posses a significant
occupancy but it does not constitute a trap. Thus the search
for the most relevant trap goes
after non-native local energy minima which have the
biggest occupancy.
In the limit of $T \rightarrow 0$,  weights associated with all other
local energy minima states become small. 

We now focus on the interpretation of the trap states to
demonstrate the qualitative validity of Figure 1.
We place the system in the trap state and ask what is
the best trajectory which leads from the trap state to the 
native state. The best trajectory is defined to be one in which
the highest energy state is lower than the highest
energy state on all other trajectories.

%%++++++++++++++++++++++++++++++++++++++++++++++++++++++++++++++++++++++++++++++
%FIGURE 6
\begin{figure}
\epsfxsize=3.2in
\centerline{\epsffile{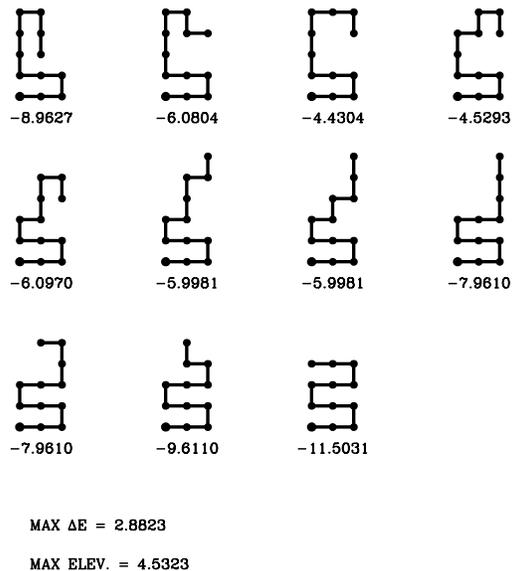}}
\caption{
The best trajectory linking the trap state of sequence A
(shown at the top left) and the native state. The energies involved
are shown below each conformation. The biggest single step 
and global energy expenses are indicated at the bottom.
}
\end{figure}

%%++++++++++++++++++++++++++++++++++++++++++++++++++++++++++++++++++++++++++++++

Figure 6 shows the trap state (of energy -8.9627) for sequence A and 
the best corresponding  trajectory  that connects it to the
native state. The trajectory reaches the 
native state in 8 steps and it elevates in energy by 4.5323, relative to the
the energy of the trap state, after accomplishing the second step.
Thus it takes only two uphil steps to enable a flow to the
native state. The first of these steps 
breaks two contacts and it costs 2.8823 in energy.
Thus this biggest single step energy cost is what determines the
value of the energy barrier in the Arrhenius law for $t_{\rm fold}$,
as derived through the Monte Carlo technique.
On the other hand, $\delta E_1$, that determines $\tau _1$, 
appears to be governed by the energy costs to make the initial
two-step climb to the point of highest elevation. In a Monte Carlo
process, two consecutive climbs up are not very likely at low $T$'s
and can be missed. However, the exactly derived longest relaxation
time indicates the true nature of the barrier. 

The kinetics of sequence B is quite similar to that of sequence A.
Thus what distinguishes between the two sequences are their
thermodynamic stabilities as measured by $T_f$. 
Figure 7 shows the steps of the best trajectory from
the the most important trap state for sequence B (of energy -8.4431)
to the native state.
The trap state is almost identical in shape
to the two native conformations: it is only the positioning of the
middle portion of the conformation that is not quite right.
And yet it takes 10 steps, like in Figure 6, to accomplish the
transition between the trap and native states.
The overall barrier that is climbed is equal to 4.4 which is
close to the
Arrhenius barrier $\delta E_1$ in $\tau _1$ of 4.0.
The biggest single step energy cost of 2.7478 on the
trajectory is, however, further away from $\delta E$ in 
$t_{fold}$ of about 3.6.
The important observation is that for the bad folder B the 
most relevant trap conformation
is still in the folding funnel of the native state.
The presence of other low lying valleys that compete
with the native valley does not really
affect the trapping effects in the native funnel. Instead it
generates low stability of the native state through a substantial
equlibrium probability of being in some other valleys.

%%++++++++++++++++++++++++++++++++++++++++++++++++++++++++++++++++++++++++++++++
%FIGURE 7
\begin{figure}
\epsfxsize=3.2in
\centerline{\epsffile{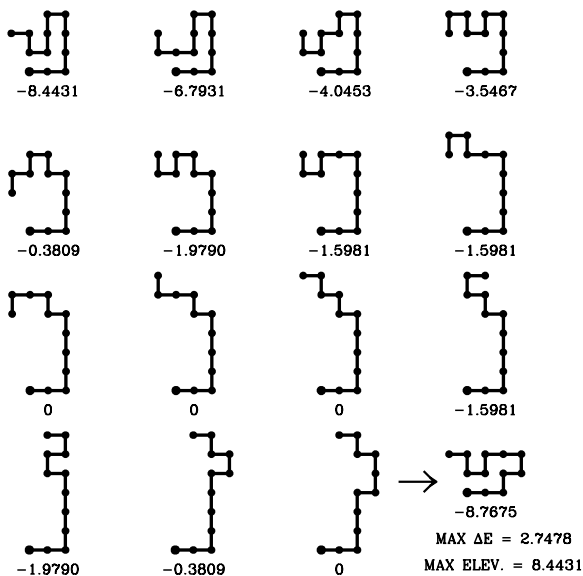}}
\caption{
As in Figure 6 but for sequence B.
}
\end{figure}
%%++++++++++++++++++++++++++++++++++++++++++++++++++++++++++++++++++++++++++++++

Even though our exact results have been obtained in a toy model of
only 12 monomers, it is expected that the essential physics of
protein folding and the distinction between the good and bad folders
are as illustrated here. Tools to study larger and off-lattice
systems in this spirit remain to be developed.

We thank Jan Karbowski for collaboration and T. X. Hoang for discussions.
This work was supported by  KBN (Grant No. 2P03B-025-13),
Polonium, CNRS-UMR 7556, NASA,
and the Applied Research Laboratory at Penn State.

\end{document}